# The Thin Gap Chambers database experience in test beam and preparations for ATLAS

Y. Benhammou, S. Bressler, E. Etzion, D. Lellouch, L. Levinson, S. Tarem[1]

*Abstract* - Thin gap chambers (TGCs) are used for the muon trigger system in the forward region of the LHC experiment ATLAS. The TGCs are expected to provide a trigger signal within 25 ns of the bunch spacing. An extensive system test of the ATLAS muon spectrometer has been performed in the H8 beam line at the CERN SPS during the last few years. A relational database was used for storing the conditions of the tests as well as the configuration of the system. This database has provided the detector control system with the information needed for configuration of the front end electronics. The database is used to assist the online operation and maintenance. The same database is used to store the non event condition and configuration parameters needed later for the offline reconstruction software. A larger scale of the database has been produced to support the whole TGC system. It integrates all the production, QA tests and assembly information. A 1/12th model of the whole TGC system is currently in use for testing the performance of this database in configuring and tracking the condition of the system. A prototype of the database was first implemented during the H8 test beams. This paper describes the database structure, its interface to other systems and its operational performance.

## I. INTRODUCTION

Large Hadron Collider (LHC) currently under construction at the European Center for Nuclear Research (CERN) is designed to generate proton-proton collision beams at the energy of 7 TeV. The ATLAS detector, built for the LHC experiment, will collect events at a proton proton bunch crossing rate of 40 MHz, between three to 23 collisions every 25 ns.

The ATLAS Muon spectrometer [1] has been designed to provide a standalone trigger on single muons with transverse momentum of several GeV, as well as to measure final state muons with a momentum resolution of about 3% over most of the expected momentum range. The ATLAS Muon spectrometer is a $4\pi$ detector consists of four types of detector technologies. Monitored Drift Tube (MDT) chambers and Cathode Strip Chambers (CSC) are used for the precision measurement of muon tracks. Resistive Plate Chambers (RPC) are employed for triggering muons in the barrel region where Thin Gap Chambers (TGC [2-3]) stations serve the same purpose in the higher background region of the end-cap.

The Muon end-cap trigger consists of seven layers of TGCs: one triplet (each consists of three wire planes) and two TGC doublets (two wire planes). The chambers are mounted on big wheels in the end-cap region about 14 m away from the interaction point. A TGC consists of a plane of closely spaced wires maintained at a positive high voltage (HV), sandwiched between resistive grounded cathode planes with a very small anode wire to cathode plane gap distance. The ATLAS TGCs are built at three production lines: Weizmann Institute of Science (Israel) building 2,160 chambers, KEK (Japan) building 1,056 chambers and Shandong University (China) building 384 chambers.

After chamber completion, the TGCs pass a thorough program of quality assurance to ensure the efficient performance of the chambers during more than ten years of operation in the LHC high rate environment [4]. This program includes a detailed mapping of the detectors response using cosmic rays hodoscopes [5-6], as well as checking the chambers behavior using a high rate radiation source. Certification tests are performed at CERN before the installation in ATLAS.

The TGC detectors will be inaccessible during operation due to the high radiation levels in the ATLAS cavern. Therefore a Detector Control System (DCS) is used to monitor important detector and environmental parameters, calibrate, set and maintain the configuration of the front end (FE) electronics, and take appropriate corrective actions to maintain detector stable and reliable performance [7]. The TGC DCS is implemented with PVSS-II, a dedicated a commercial - Supervisory Control And Data Acquisition (SCADA) product. To verify the TGC system performance, a set of tests was performed under muon beams at the CERN H8 testing area in the years 2000-2004. A vertical slice of the full trigger and data acquisition electronics and fully instrumented TGCs (two doublets and a triplet) were placed in the beam to form a trigger "tower". These tests confirmed the reliable operation and high efficiency of the TGC trigger. The TGC Data Acquisition (DAQ) and DCS developers have used the H8 test-beam to demonstrate prototype readout, configuration and calibration methods. The DCS controlled the power supply to the chambers as well as recorded the environmental temperature using dedicated sensors.

The complex structure of the system, with different

[1] Manuscript received June 16 2005. This work is supported by the Israel Science Foundation and the German Israeli Foundation.

Y. Benhammou and E. Etzion are with the School of Physics and Astronomy, Raymond and Beverly Sackler Faculty of Exact Sciences, Tel Aviv University, Tel Aviv 69978, Israel (telephone: +97236408303, e-mail: ybenham@lep1.tau.ac.il).

S. Bressler and S. Tarem are with Physics Department, Technion, Haifa, Israel.

D. Lellouch and L. Levinson are with the Particles Physics Department, Weizmann Institute of Science, Rehovot 76100, Israel.

component hierarchies for chambers, trigger, read out, and DCS, requires a relational database to describe the various interconnections between the system components (see fig 1.). For instance, one electronic module needs to be associated to the chamber from which it gets its signals, to the trigger coincidences to which it contributes, and again to the control path by which it is configured. Each connection might go through a very different path but we should be able to express all these different connection and access paths and hierarchies in a natural and efficient way.

## II. DATABASE STRUCTURE

### A. Database goals

The database is aimed at storing the non event data needed for the detectors operations and maintenance, the online readout and the offline analysis.
The database is required to fulfill the following requirements:
- record non event data,
- provide data for efficient detector configuration via the DCS
- provide data for efficient detector configuration via the DAQ,
- assist in detector calibration and diagnosis,
- provide data to the offline programs,
- provide data to the online programs,
- support for tagging Interval of Validity (IoV) and versions for the stored data.

Key features are the ability to navigate throught the different hierarchies in which the various chambers and electronics modules can be viewed. For example, an ASIC located on a chamber, is accessed by the DCS via a CAN node (DCS-PS) on the CANbus path. The same ASIC is accessed by the DAQ via a special board (StarSwich). In case of problem, the DAQ should be able to know which electronic boards are involved namely to find the appropriate ASIC and then to return to the DCS paths and locate the relevant board.

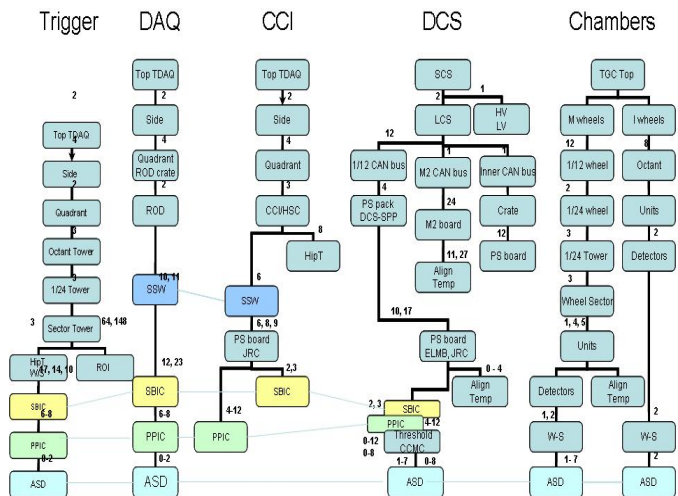

Fig 1: schematic views and relations between the main different paths: DAQ, TRIGGER, DCS, Crate Controller Interface and Construction

For that purpose, several methods for data storage were considered: flat files and structured directories solution, simple database with large object (BLOB), object-oriented (OO) database and relational database. While retrieval structured data (parsing) in the two first options are really time consuming, the last two could fit well our needs. The choice of relational database was natural due to the fact that our DCS software platform, PVSS, is natively implemented to work with such a database type.

### B. Database design

The design of the relational database has been created according to an OO schema. Fig. 2 depicts the general table structure of our implementation. The table DEVICE contains the general detector elements: ASIC, CHAMBER (each with its own unique ID). The ARGUMENT table contains the parameters of these elements linked with the previous table with the unique ID of the elements. The PRODUCT table allows creating the instances of these devices, again linked to them with their unique ID. This scheme allows to easily add a new parameter to a device or to install a new product. This is done by adding one row to the appropriate table.

All the details of the configuration (name, time, date, etc…) are saved in the CONFIGURATION table where each entry carries its own id. In the CONFDB table, there are four important fields: product id, argument id, configuration id and value. The three first are references to the corresponding tables and the last one is the value of the parameter to set. The read values are recorded in the condition table (CONDDB) which is identical to the CONFDB table, the values are a snapshot of values imported from the DCS and DAQ. The DCS doesn't use tables from CONDDB to configure the hardware, but values may be imported from CONDDB to CONFDB for later use in configuration.
In order to navigate between the different paths, a HIERARCHY table has been created. This table contains three main fields: the product id "father", the product id "son" and the path name. This allows a product to be either the "son" or the "father" in different hierarchies.
This scheme makes it easy to navigate through complicated structures via simple SQL queries.

## III. TEST BEAM

### A. Set up

A large scale test stand of the ATLAS detector including all the Muon spectrometer components has been operated in the CERN north area on the H8 beam line during the years 2000-2004 [8]. A beam of up to 320 GeV provided by the SPS accelerator was used to study different aspects of the spectrometer. In the end-cap stand which reproduced a muon spectrometer end-cap sector there were 11 MDT chambers fully instrumented with FE and read out through one Muon Readout Driver (MROD). The chambers were equipped with the complete alignment system and calibrated sensors for absolute alignment. For triggering there were seven TGC stations (two doublets and one triplet) equipped with two types of FE boards with ten integrated circuits. This represents a set of 350 settings parameters and another 200 readout parameters.

The DCS controlled one high voltage board, collected charge histograms from each chamber, set and monitored the thresholds and the ASIC programmable parameters. The DAQ and the trigger were in charge of 20 programmable ASICs. The database has been implemented using MySQL which was a simple and free relational database.

### B. Slow Control

The TGC DCS system was used to control the CAEN LV HV power supplies by the PVSS process communicating via an OPC server [9]. The DCS was used to set the readout threshold. The DCS was running a chamber charge monitoring embedded in the DCS boards. The DCS board provided monitoring of the analog charge of a wire in a chamber over some time interval.

The database communicates directly with the DCS system via the PVSS connectors. These connectors allow the DCS to directly store and retrieve the relevant conditions and configuration parameters. Two types of functionalities have been integrated into the PVSS code: "write" (to write from the DCS to the database) and "read" (to extract value from the database to the DCS which uses them to set a configuration).
The "write" and "read" functionalities require three arguments: product, its parameter and configuration name. It is possible to read/write a subset of parameters/products or go through a full set of the configuration data.
A GUI of the database operation has been integrated with the general DCS operation panels.
During the run, automatic parallel threads launched by PVSS monitored the settings and recorded values such as temperature and high voltage in the database with different time intervals, or as a response to external trigger. Chamber charge histograms collected by the DCS were also stored to the database.

### C. Data Acquisition

At the beginning and the end of each run, the DAQ stores a set of operation parameters such as run number, software release etc... During the first beam tests, the DAQ wrote a flat file containing all the relevant parameters. A triggered PERL script parsed the file and wrote the parameters into the database. This was later modified to a direct connection between the DB and the DAQ C++ code. Few classes have been created to map and retrieve data from the database. These classes are working as an intermediate factory layer between SQL and C++.

### D. Results

All the above features have been implemented in a flexible, easy to use and maintain MySQL design. The connectivity to DCS and DAQ evolved along the period of the test beam reaching full integrated system using direct access without an intermediate API layer. The application allowed histograming and monitoring using dedicated tools (ROOT, Microsoft tools) and web access capabilities (JAVA). The database performance was successfully demonstrated during several operation periods in the H8 test beam.

### IV. DETECTOR SECTOR TEST

The construction of the first sector which corresponds to $1/12^{th}$ of one of the six TGC wheels started at the beginning of the year at CERN. Before mounting, the TGCs have been certified going through rigorous tests starting at the production sites and finalizing at CERN. In order to simplify integration and later the data access, the design of the certification database which tracks the results of these tests is based on the same structure (unique ID) as the test beam implementation.

Going from test beam scale to ATLAS scale we had to slightly modify the database scheme. The methods must evolve from reading out, controlling and calibrating seven chambers, two electronics boards and about 500 parameters, to 3,600 chambers and 1,500 electronics boards with about 300,000 parameters.
The first step in this migration was to test the performance of data recording and system configuration for 1/12th of a TGC wheel.
A DCS setup was established in Israel to test the DCS performance and long term robustness. This setup represents a full DCS CAN bus branch, and includes 54 electronics boards. It can control 500 integrated circuits and 12,500 parameters to set and read. This "Branch Test" was also used to test performance issues related to DB-DCS communications. We found that, to improve the configuration performances, some data organization tasks had to be shifted from PVSS to the database and the database design had to be modified accordingly. In order to shift code to the database (utilization of database trigger functionality) and to ameliorate performances (using bind variables) we moved from MySQL to ORACLE 9i which is accessible and maintained by the CERN IT group.
To improve the database design, the CONFDB table has been modified to more closely model the physical design of the configured electronics. The table was split to four different tables which results in a better data granularity and thus a faster response (fig 2.).

Preliminary tests of the modified scheme show significant improvement in the performance of the connection between the DB and the DCS.

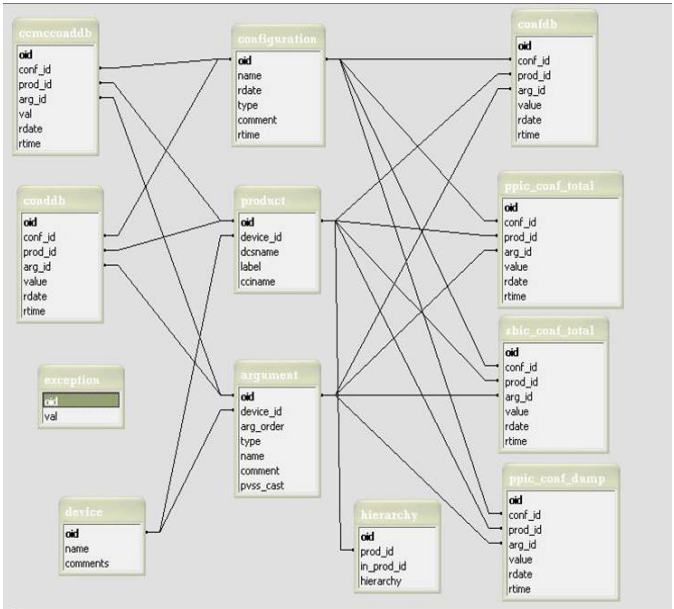

Fig 2 : database design has been updated after the test beam to ameliorate the performances and be able to set 12,500 parameters in a reasonable time.

## V. DATABASES RELATION

The new ORACLE database combines the information recorded during the chambers production obtained from the different production sites. Information from the three production sites (China, Israel and Japan) has been exported from the variety of storage technologies (ACCESS, MySQL, MSQL and object oriented DB). This includes the detailed quality assurance procedures realized in the cosmic test stands and at CERN. The database is used to assist and guide the assembly and construction of the TGC system. The data stored during the installation of the chambers and the electronics as well as the chambers performance in the quality control tests will supply the initial configuration parameters. Figure 3 represent the global structure of the unified non-event TGC DB.

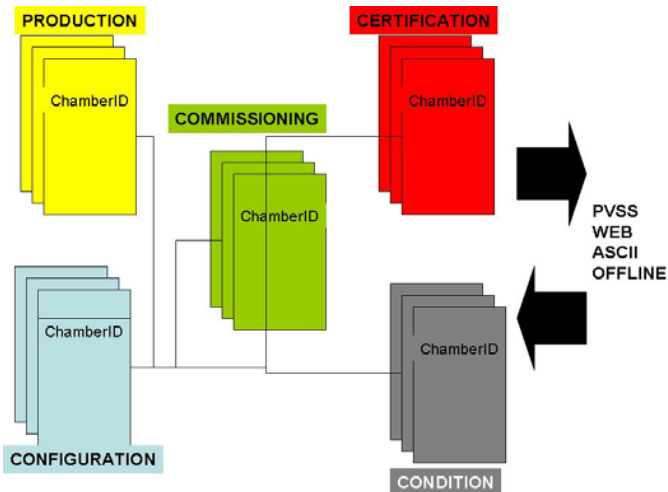

Fig 3: relation between the tables from different steps in the case of a chamber. Because of the unique chamber id, it is easy to navigate between the difference component (such as production and conditions inofrmation).

## VI. SUMMARY

A relational database has been designed to store the TGC running conditions provided by both DCS and DAQ as well as configuration parameters needed for the FE and operation. This scheme has been primarily tested in the H8 beam line at the CERN. The DB design has to take into account the specific problems of hierarchies. Indeed, different clients (DCS, DAQ, TRIGGER etc…) have to be able to access the same parameters (stored in the database) trough different paths (e.g. ASIC, boards...). The database model has been designed to represent all the electronic devices in a table. All objects (instance of the device and physical object) are stored in a different table and all the parameters of each device are located in a separate table. The one-to-N relations are represented by the "references constraint" between the device and the objects or the device and the arguments. The first design has been implemented with MySQL and validated in the different test beam representing about 0.2% of the parameters space.

To support a larger number of parameters needed for the configuration of a sector of a TGC wheel, the design of the database has been slightly modified and transferred to ORACLE. This also allowed shifting software from the DCS to the database. The database now also integrates the production, quality assurance test and assembly information in order to facilitate the navigation between data in case of problems.

ACKNOWLEDGMENT

We thank all our colleagues from the TGC groups in China, Japan and Israel.